\documentclass[aps,prl,showpacs,twocolumn]{revtex4}
\usepackage{latexsym}
\usepackage{amssymb}
\usepackage{amsmath}
\usepackage[dvips, pdftex]{graphicx}
\usepackage{tikz}
\newcommand{\ket}[1]{\left| #1 \right>} 
\newcommand{\bra}[1]{\left< #1 \right|} 
\newcommand{\meanv}[1]{\left< #1 \vphantom{#1} \right>} 
 
\newcommand{\pms}{\mathbin{\tikz[x=1.4ex,y=1.4ex,line width=.1ex] \draw (0.0,0) -- (1.0,0) (0.5,0.08) -- (0.5,0.92) (0.0,0.5) -- (1.0,0.5);}}%
\newcommand{\p}{\mathbin{\tikz[x=1.4ex,y=1.4ex,line width=.1ex] \draw (0.0,0) -- (0.0,0) (0.5,0.08) -- (0.5,0.92) (0.0,0.5) -- (1.0,0.5);}}%
\newcommand{\m}{\mathbin{\tikz[x=1.4ex,y=1.4ex,line width=.1ex] \draw (0.0,0) -- (1.0,0) ;}}
\usepackage{color}
\definecolor{red}{rgb}{1.0,0.0,0.0}
\definecolor{blue}{rgb}{0.0,0.0,1}
\definecolor{green}{rgb}{0.29, 0.33, 0.13}
\begin{document}
\title{Explanation of the intriguing quantum phenomenon of the off-resonant cavity mode emission}
\author{Santiago Echeverri-Arteaga$^{1}$}
\author{Herbert Vinck-Posada$^{1}$}
\affiliation{$^{1}$Departamento de F\'isica, Universidad Nacional de Colombia, 111321, Bogot\'a, Colombia}
\author{Edgar A. G\'omez$^{2}$}
\affiliation{$^{2}$Programa de F\'isica, Universidad del Quind\'io, 630004, Armenia, Colombia}
\begin{abstract}
We theoretically investigate the unexpected occurrence of an extra emission peak that has been experimentally observed in off-resonant studies of cavity QED systems. Our results within the Markovian master equation approach successfully explain why the central peak arises, and how it reveals that the system is suffering a dynamical phase transition induced by the phonon-mediated coupling. Our findings are in perfect agreement with previous reported experimental results and for the first time the fundamental physics behind this quantum phenomenon is understood.
\end{abstract}
\maketitle
\begin{figure}[h!]
\includegraphics[scale=1]{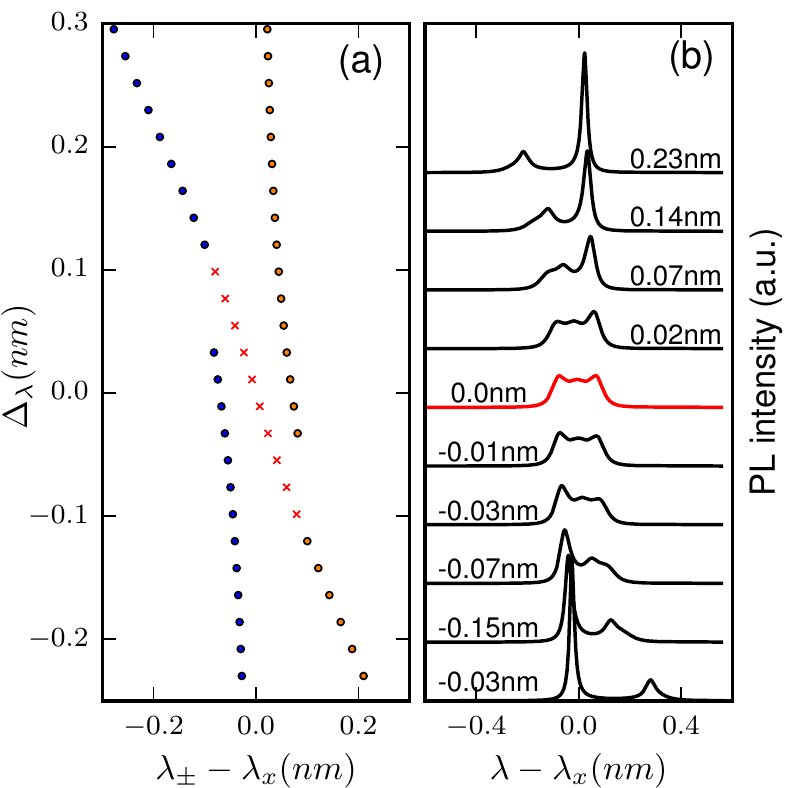}
\caption{(color online) (a) Spectral peak positions of the polaritons $\lambda_\pm$ together with the position of the central peak for various detunings $\Delta_\lambda$. (b) Emission spectra for various detunings $\Delta_\lambda$ showing how the central peak grows as the QD enters in resonance, in perfect agreement with the experimental data reported by Hennessy~{\it et al.}. The numerical simulations were performed using the parameters: $\omega_c=1309.78 meV$, $g=0.12 meV$, $\kappa=0.032 meV$, $\gamma_x=0.0112 meV$, $\gamma_{\theta}=0.17 meV$ and $P_x=0.004 meV$.}\label{fig-henessy}
\end{figure}

\textit{Introduction.}\textendash
During the last decades an intriguing quantum phenomenon in the cavity quantum electrodynamics systems (cQEDs) has been observed.
More precisely, the photoluminescence (PL) measurements of quantum dots (QDs) coupled to semiconductor cavities under off-resonant QD-cavity coupling have shown that an unexpected off-resonant PL emission of the cavity mode can be observed as a function of exciton-cavity detuning. This fact has been identified and reported by several experimental groups, and particularly some researchers have argued that this phenomenon could be the result of the background emitters, as well as the contribution of a variety of mechanisms including photons and acoustic phonons to the cavity emission~\cite{Press:2007}. Furthermore, some works related with off-resonant experiments in biexitonic systems have shown an off-resonant cavity emission which has been attributed to a manifestation of off-resonant cavity feeding~\cite{Winger:2008}. Moreover, strong experimental evidence supports that an intrinsic off-resonant coupling mechanism between a QD and the cavity mode exists~\cite{Kaniber:2008}. Within the experimental scenario involving off-resonant cQED systems, Hennessy {\it et al.}~\cite{Hennessy:2007} performed an extraordinary experiment in the strong-coupling regime with exceptional results, where the emergence of an additional peak at the cavity frequency was clearly observed. However, this peak could not be explained successfully by them. More recently, a similar experiment confirm the spectral triplet at the low excitation power regime~\cite{Ota:2009}. Theoretical studies have been performed for understanding this quantum phenomenon as well as the fundamental physics behind it \cite{Yamaguchi:2008,Naesby:2008,Hughes:2009}, and its origin has been attributed to mechanisms as pure QD dephasing~\cite{Auffeves:2009,Gonzalez:2010},  non-markovian processes~\cite{Yamaguchi:2009}, electron-phonon interactions \cite{Hohenester:2009,Hohenester:2010,Hughes:2011,Roy:2011b} as well as effects due to multiexciton complexes \cite{Winger:2009} or Coulombian interactions between charged QDs \cite{Chauvin:2009}. Despite the efforts of many researches this quantum phenomenon still remains unexplained and surprisingly, in the last years, this topic apparently seems forgotten. In this Letter, we unequivocally demonstrate that the spectral triplet which has been observed experimentally is a consequence of a dynamical phase transition in the system.
\\
\textit{Theoretical model.}\textendash
Since we are interested in modeling a solid state optical system based on a photonic-crystal cavity with an embedded single QD under off-resonant QD-cavity coupling, we consider the most simple and accurate quantum model for describing the interaction between radiation and matter. It is, the Jaynes-Cummings (JC) model that, in the dipole and rotating wave approximations, reads ($\hbar=1$) $\hat{H}=\omega_{x}\hat{\sigma}^{\dag}\hat{\sigma}+\omega_c\hat{a}^{\dag}\hat{a}+g(\hat{a}^\dag\hat{\sigma}+\hat{a}\hat{\sigma}^\dag)$ with $g$ the dipole coupling constant between the cavity mode and the QD as a two-level system. The operator $\hat{a}$ annihilates a photon from the cavity and $\hat{\sigma}=\ket{0}\bra{1}$ is the lowering operator from the excited state $\ket{1}$ to the ground state $\ket{0}$, moreover the detuning between the QD and the cavity mode is given by $\Delta=\omega_{x}-\omega_c$ with $\omega_{c}$ and $\omega_{x}$ being the corresponding frequencies associated to the cavity and the QD exciton transition, respectively. In order to get a rich phenomenology in our model, we have incorporated different decoherent system-environment processes via the Markovian master equation in the Lindblad form:
\begin{eqnarray}\label{eq1}
\frac{d\hat{\rho}}{dt}&=&-i[\hat{H},\hat{\rho}]+\frac{\kappa}{2}\mathcal{L}_{a}(\hat{\rho})+\frac{\gamma_{x}}{2}\mathcal{L}_{\hat{\sigma}}(\hat{\rho})+\frac{P_{x}}{2}\mathcal{L}_{\hat{\sigma}^{\dagger}}(\hat{\rho})\notag \\ &+&\frac{\gamma_{\theta}}{2}\mathcal{L}_{\hat{\sigma}^{\dagger}\hat{a}}(\hat{\rho}),
\end{eqnarray}
where $\mathcal{L}_{\hat{X}}$ is the superoperator that is defined, for a generic operator $\hat{X}$, as $\mathcal{L}_{\hat{X}}(\hat{\rho})=2\hat{X}\hat{\rho}\hat{X}^{\dagger}-\hat{X}^{\dagger}\hat{X}\hat{\rho}-\hat{\rho}\hat{X}^{\dagger}\hat{X}$. Furthermore, the irreversible processes induced by the environment on the system are the leakage of photons from the cavity at rate $\kappa$, the spontaneous emission $\gamma_x$ and the incoherent QD pumping $P_x$. It should be emphasized that only at low temperature the pure dephasing mechanism can be safely neglected. Here we have taken into account this fact since the experimental conditions achieved by Hennessy {\it et al.} involve low temperatures of order of $4K$. It is worth mentioning that the dephasing mechanism fails to explain the off-resonant cavity mode emission as is well reported in the literature \cite{Hughes:2009,Yamaguchi:2008,Suffczyński:2009}. Finally, the last term in the master equation describes the off-resonant QD-cavity coupling proposed by Majumdar {\it et al.} \cite{Majumdar:2010} few years ago, as a physical mechanism able to compensate the QD-cavity frequency difference by the creation (or annihilation) of phonons at decay rate $\gamma_\theta$. Particularly, this phonon-mediated coupling mechanism has recently inspired investigations on electron-acoustic-phonon scattering in QD-cavity systems \cite{Roy:2011}, as well as effects due to the phonon scattering on exciton-cavity interaction in the weak-coupling regime \cite{Jarlov:2016}. We compute numerically the emission spectra of the coupled system, according to the Wiener-Kinchine theorem, through the Fourier transformation of the cavity field autocorrelation  $S(\omega)=\int_{-\infty}^{\infty}\meanv{\hat{a}^{\dagger}(\tau)\hat{a}(0)}e^{-i\omega \tau}d\tau$. 
To calculate the two-time correlation function and subsequently the emission spectrum, we use the well-known quantum regression formula (QRF)~\cite{Perea:2004}. In particular, our numerical simulations have been conducted for comparison purposes with the experimental results shown in Fig.~$3$ from Ref.~\cite{Hennessy:2007}. Further, wherever possible the parameters chosen for the simulations were taken from the experimental setup used in this reference. In Fig.~\ref{fig-henessy}(a) are shown the peak positions extracted from the emission spectra for various detunings $\Delta_{\lambda}=\lambda_x-\lambda_c$, and they are plotted in blue and yellow dots. Additionally, the red cross marks show clearly the position of the central peak evidencing its existence. It is shown in Fig.~\ref{fig-henessy}(b) that the emission spectra for the two anticrossing polariton states near zero detuning and the central peak arising in perfect agreement with the experimental evidence reported. Notice that the resonant case is shown as a solid red line and for other cases of $\Delta_{\lambda}$ the emission spectra are shown as solid black lines. It is worth mentioning that experiments with \textbf{Ba} atoms strongly coupled to a high-finesse cavity have shown similar triplets \cite{Childs:1996}, which could be explained in terms of this theoretical approach by assuming that the phonon-mediated coupling comes from the radiation pressure.
To the best of our knowledge, there is not any other theoretical work based on Markovian master equation that shows a perfect agreement with the experimental results obtained by Hennessy {\it et al.}, as well as any other that clearly explains the emergence of this central peak. In fact, the fundamental physics related to this quantum phenomenon has not  been understood yet and it remains as an open problem. Motivated by the above and recognizing that the master equation given by the Eq.~(\ref{eq1}) cannot be solved for an arbitrary set of parameter values in a closed form, we turn our attention to the study of the emergence of the central peak within the framework of a master equation without gain~\cite{Torres:2014}.
\\
\textit{Dynamical phase transition}\textendash 
It is well-known that the JC Hamiltonian has the conserved quantity $\hat{N}_{exc}=\hat{a}^{\dagger}\hat{a}+\hat{\sigma}^{\dagger}\hat{\sigma}$ known as the excitation-number operator (where its eigenvalues define the rungs in the JC ladder), which is diagonal in the bare-states basis $\big\{\ket{n-\alpha,\alpha}\equiv\ket{n}\vert_{n=0}^{\infty}\otimes\ket{\alpha}\vert_{\alpha=0}^{1}\big\}$. In this basis, $n$ and $\alpha$ represents the number of photons in the cavity and one of the two possibles states of the QD, respectively. It is noticeable that $[\mathcal{L}_{\hat{\sigma}^{\dagger}\hat{a}},\hat{N}_{exc}]=0$, implies that the phonon-mediated coupling preserves the number of excitations in the system. Moreover, by taking into account that this consequence remains also valid for the non-Hermitian operator $\hat{K}=\hat{H}-i\gamma_{x}\frac{1}{2}\hat{\sigma}^{\dagger}\hat{\sigma}-i\kappa \frac{1}{2}\hat{a}^{\dagger}\hat{a}$, we can construct a Lindblad master equation without  gain~\cite{Torres:2014}, {\it i.e} $P_x=0$, as follows $d\hat{\tilde{\rho}}/dt=-i[\hat{K},\hat{\tilde{\rho}}\,]+\gamma_{\theta}\mathcal{L}_{\hat{\sigma}^{\dagger}\hat{a}}(\hat{\tilde{\rho}}\,)\equiv{\cal L}\hat{\tilde{\rho}}$. With this approach we are able to investigate how the phonon-mediated coupling is the responsible for the occurrence of the central peak, it without a strong influence of the dissipative terms at rates $\kappa$ and $\gamma_x$. A remarkable aspect of this master equation is that the full Liouvillian $\mathcal{L}$ also preserves the number of excitations, and therefore it can be partitioned into subspaces, more precisely in $4-$dimensional subspaces $\mathcal{L}^{n,m}$ that can be written in terms of the $4\times4$ matrices. Moreover, each subspace is spanned by a pair of excitations $n$ and $m$ and has associated the eigenvalue problem $\mathcal{L}^{n,m}{\mathbf U}^{n,m}=\lambda^{n,m}{\mathbf U}^{n,m}$. For this particular partitioning of the subspaces, it should be taken into account that an arbitrary density matrix can be written as $\hat{\tilde{\rho}}=\sum_{n,m,\alpha,\beta}\varrho_{n,m}^{\alpha,\beta}\ket{n-\alpha,\alpha}\bra{m-\beta,\beta}\equiv\sum_{n,m}\rho_{n,m}$, where the last term can be associated to a $2\times2$ matrix with its elements given by $\varrho_{n,m}^{\alpha,\beta}$. Moreover,  it is straightforward to prove that $[\hat{N}_{exc},\rho_{n,m}]=(n-m)\rho_{n,m}$ implies that the matrix $\rho_{n,m}$ has a definite number of $n$ and $m$ excitations mentioned above. Interestingly, the $n$th subspace $\mathcal{L}^{n,n-1}$ is spanned by the operators: $\hat{a}^{\dagger}_{n,0}=\ket{n,0}\bra{n-1,0}$, $\hat{a}^{\dagger}_{n,1}=\ket{n-1,1}\bra{n-2,1}$, $\hat{\sigma}^{\dagger}_{n}=\ket{n-1,1}\bra{n-1,0}$ and $\hat{\zeta}^{\dagger}_{n}=\ket{n,0}\bra{n-2,1}$ which correspond to the set of four operators required by the QRF for finding the emission properties in the system. Therefore, it is expected that any significant information about the emission properties of the system comes from this subspace. It explicitly is given by 
\begin{equation}
\mathcal{L}^{n,n-1}=
\begin{bmatrix}
\frac{\kappa}{2} & -i\Omega_{n-1} & i\Omega_{n} & \sqrt{n(n-1)}\gamma_{\theta} \\
-i\Omega_{n-1} & \frac{\gamma_x-(n-1)\gamma_{\theta}}{2} & 0 & i\Omega_{n}   \\
i\Omega_{n} & 0 & \frac{2\kappa-\gamma_x-n\gamma_{\theta}}{2} &-i\Omega_{n-1}   \\
    0 & i\Omega_{n} & -i\Omega_{n-1} & \frac{\kappa-(2n-1)\gamma_{\theta}}{2}
\end{bmatrix}
\end{equation}
where we have defined the vacuum Rabi frequency as $\Omega_n=g\sqrt{n}$ and $\Delta=0$. The eigenvalues $\lambda^{n,n-1}_{\pms\,\pms}$ of this subspace are directly related to the eigenfrequencies  $\omega^{n,n-1}_{\pms\,\pms}=\text{Im}[\lambda^{n,n-1}_{\pms\,\pms}]$ (position peaks) and the linewidths $\Gamma^{n,n-1}_{\pms\,\pms}=\text{Re}[\lambda^{n,n-1}_{\pms\,\pms}]$ of the emission spectrum. This fact accounts for the four possible transitions between the dressed states in the JC rungs $n$ and $n-1$, as is expected in the case of $\gamma_\theta=0$. But, for the case $\gamma_\theta\neq0$ a huge broadening in two linewidths $\Gamma^{n,n-1}_{\p\,\pms}$ is found in our numerical results, hence their corresponding transitions do not contribute significantly to the emission spectrum.
\begin{figure}[h!]
\centering
\includegraphics[scale=1.]{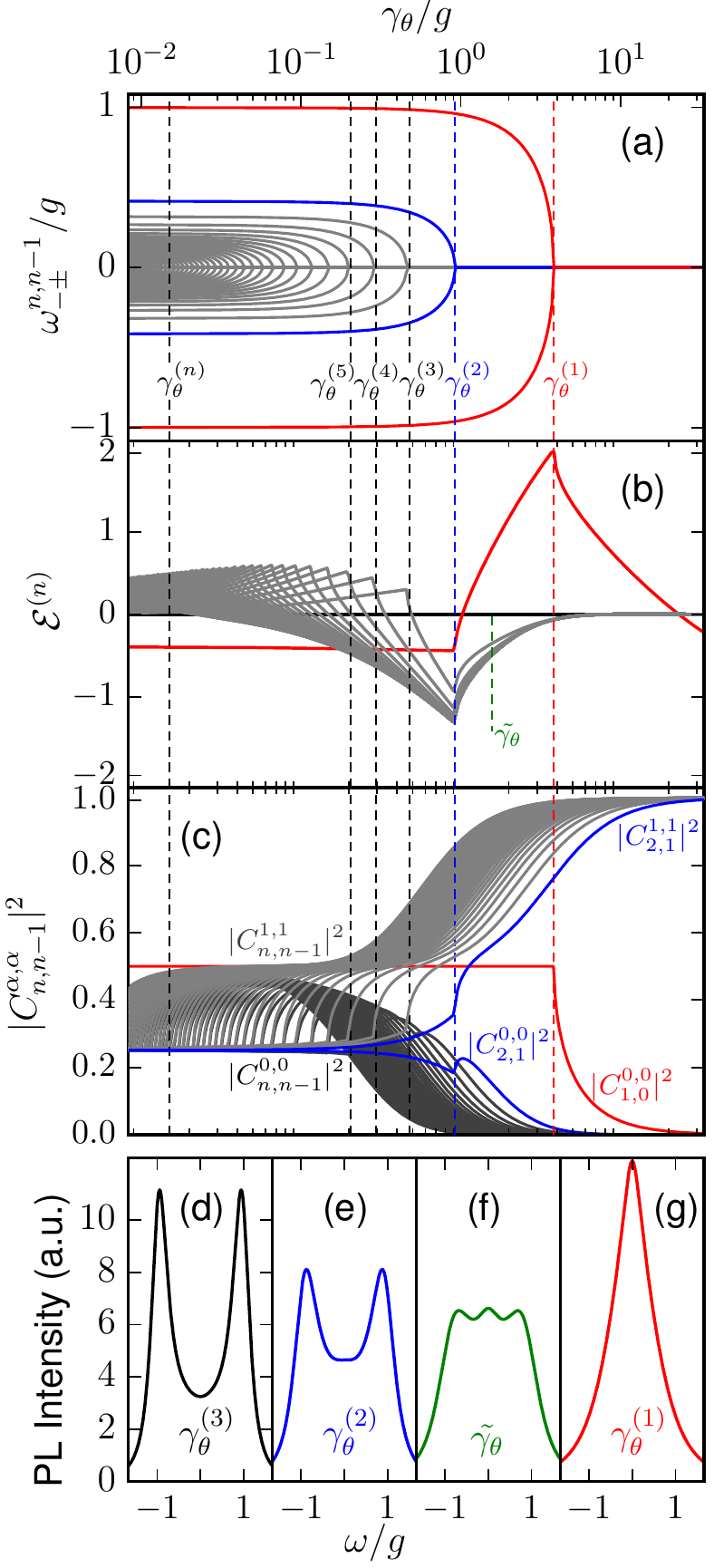}
\caption{(color online) Numerical results based on master equation approach without gain for
(a) $\omega^{n,n-1}_{\protect\m\,\protect\pms}/g$, (b) $\mathcal{E}^{(n)}$ and (c) $|C^{\alpha,\alpha}_{n,n-1}|^2$  with $\alpha=0,1$, as a function of $\gamma_{\theta}/g$. Some critical values are shown as vertical dashed lines at $\gamma_{\theta}^{(1)}/g=3.79$, $\gamma_{\theta}^{(2)}/g=0.92$,  
$\gamma_{\theta}^{(3)}/g=0.48$, $\gamma_{\theta}^{(4)}/g=0.30$, $\gamma_{\theta}^{(5)}/g=0.21$, and any other critical value can be found through the approximation at $\gamma_{\theta}^{(n)}$ (see text for details). Four emission spectra are shown in (d)-(g) at   $\gamma_{\theta}^{(3)}$, $\gamma_{\theta}^{(2)}$,
$\tilde{\gamma}_{\theta}=1.57$ and $\gamma_{\theta}^{(1)}$, respectively. Additionally, the
parameters used for the numerical calculations are found in the caption to Fig.~\ref{fig-henessy}.}\label{fig-1}
\end{figure}
In contrast, the eigenfrequencies $\omega^{n,n-1}_{\m\,\pms}$ and their corresponding linewidths will play an important role in the emission properties. In Fig.~\ref{fig-1}(a) these eigenfrequencies are shown as a function of $\gamma_{\theta}$ for the first fifty JC rungs, particularly the first and second rungs are depicted as solid red and blue lines, respectively. Any other eigenfrequency is shown as solid-grey line, moreover these eigenfrequencies merge to the cavity frequency at critical values $\gamma^{(n)}_{\theta}\approx 4g\sqrt{\frac{(4n_{1}^3+16n_{1}^{2}+10n_{1}+6)^\frac{1}{2}-(2n^3-3n^2+n)-\frac{(\kappa-\gamma_x)}{n(n+1)}}{15n_{1}^{2}+10n_{1}+6}}$ with $n_1=n(n-1)$ for $n=2,3,\dots$ and $\gamma_\theta^{(1)}=4g-(\kappa-\gamma)$. These critical values are exceptional points (EPs) since they are related with the coalescence of two eigenvalues $\lambda^{n,n-1}_{\m\,\pms}$ and the dimension of the subspace drops by one~\cite{Rotter:2010}. Here some critical values are shown as vertical dashed lines for guide eye. Before $n$th EP a simple proportionality relation between the linewidths is given by 
\begin{equation}\label{eq3}
(\Gamma^{n,n-1}_{\m\,\pms}+\Gamma^{n-2,n-3}_{\m\,\pms})/\Gamma^{n-1,n-2}_{\m\,\pms}=2,
\end{equation}
where the phenomenology of the Rabi doublets holds for three consecutive rungs as in the well-known structure of  the dissipative JC ladder. This scenario changes substantially after the EP where a bifurcation is leading to large linewidth $\Gamma^{n,n-1}_{\m\,\p}$
and gives origin to a singlet at the cavity frequency with linewidth $\Gamma^{n,n-1}_{\m\,\m}$. In other words, when the Eq.~(\ref{eq3}) is not valid an structural change occurs in the emission properties of the $n$th subspace. This fact can be analyzed in a global manner by taking into account all subspaces  $\mathcal{L}^{n,n-1}$, more precisely, the Eq.~(\ref{eq3}) handled recursively can be cast as a function $\mathcal{G}(\gamma_{\theta})=\sum_{n=1}^{\infty}\frac{\Gamma^{n,n-1}_{\m\,\m}}{2^n\Gamma^{2,1}_{\m\,\m}}$ that incorporates the effect of $\gamma_{\theta}$ on the whole space of optical transitions of the system. Thereby, any sudden change in the spectral structure of the system will come up accompanied by a non-analyticity in $\frac{d\mathcal{G}}{d\gamma_{\theta}}=\sum_{n=1}^{\infty}
\frac{\Gamma^{n,n-1}_{\m\,\m}}{2^n\Gamma^{2,1}_{\m\,\m}}\frac{\partial \mathcal{E}^{(n)}}{\partial \gamma_{\theta}}$
where is defined the dimensionless parameter $\mathcal{E}^{(n)}=\ln\frac{\Gamma^{n,n-1}_{\m\,\m}}{\Gamma^{2,1}_{\m\,\m}}$ for $n\neq2$. Interestingly, this parameter reveals how the formation of a collective state with small width takes place in the system, which is environmentally induced by the phonon reservoir.
More precisely, at $\gamma_{\theta}^{(2)}$ a resonance state is entirely created as the result of the successive overlapping of the singlet that arises once each of the EPs occurs. Additionally, this resonance state is characterized by the fact that all the singlets that contribute to its formation have reached its narrowest linewidth. It is observed in each of the $\mathcal{E}^{(n)}$ shown in Fig.~\ref{fig-1}(b) as solid-grey lines.
This resonance trapping phenomenon \cite{Jung:1999} together with the distribution of the EPs over a finite range of the parameter $\gamma_{\theta}$, implies that a dynamical phase transition (DPT) is occurring in the system~\cite{Rotter:2010}, now all the singlets that contribute to the resonance state will have almost the same linewidths at  $\gamma_{\theta}^{(1)}$.
Additionally, the parameter $\mathcal{E}^{(1)}$ shows a remarkable aspect as is depicted as a solid-red line, it remains constant and takes a negative value until the resonance state gets formed, giving it a clear evidence that the Rabi doublet from linear JC model is preserved. Nevertheless, this doublet begins to merge and rapidly enters to reinforce the smooth background emission. In order to elucidate the differences in the two emission phases in the system and the influence of the collective phenomenon, we turn our attention to the expansion coefficients $C^{\alpha,\beta}_{n,n-1}$ of the eigenvector 
${\mathbf U}^{n,n-1}_{\m\,\m}=\sum_{\alpha,\beta}C^{\alpha,\beta}_{n,n-1}\ket{n-\alpha,\alpha}\bra{n-1-\beta,\beta}$. More specifically, to the coefficients $C^{0,0}_{n,n-1}$ and $C^{1,1}_{n,n-1}$ that are related to the optical transitions where the QD is in the ground ($\hat{a}^{\dagger}_{n,0}$) and the excited state ($\hat{a}^{\dagger}_{n,1}$), respectively. In Fig.~\ref{fig-1}(c) are shown the square modulus of these coefficients and different emission properties or phases of the system can be characterized below and beyond the DPT. One phase is identified by the typical Rabi doublet associated to the transitions from the first rung of the linear (dissipative) JC model.
It is confirmed by the coefficient $C^{0,0}_{1,0}$ that remains constant until $\gamma_{\theta}^{(1)}$, where the transition rapidly gets suppressed as shown in solid-red line. The other phase brings out a decoupling of light from matter and it is recognized by a singlet at the cavity frequency since all coefficients $C^{1,1}_{n,n-1}\sim1$ and $C^{0,0}_{n,n-1}\sim0$. Particularly, this phase begins properly at $\gamma_{\theta}^{(2)}$ where the emission of the system as a whole changes substantially, this due to the restructuring of the JC ladder that is evidenced in all emission properties as a collective behavior. An interesting phenomenological scenario occurs in the region $\gamma_{\theta}^{(2)}<\gamma_{\theta}<\gamma_{\theta}^{(1)}$ where a coexistence for both phases is achieved, hence the system can operate simultaneously under two different emission regimes: the {\it doublet phase} corresponds to transitions when the QD is in the ground state and it is observed a strong coupling light-matter regime. The {\it singlet phase} involves all optical transitions when the QD is in the excited state and a pure photonic state of the cavity is observed as a signature of the decoupled regimen mentioned above. This scenario is corroborated in the emission spectrum of the system, when it is considered the master equation approach given by Eq.~(\ref{eq1}). The numerical results for the emission spectra are shown at four particular values of $\gamma_{\theta}$. The {\it doublet phase} is shown in Fig.~\ref{fig-1}(d) where clearly the Rabi splitting is  $2g$. The emission spectra in Fig.~\ref{fig-1}(e)-(f) show the coexistence region where a plateau appears indicating that the resonance trapping phenomenon has occurred and the resonance state is created by the system as a whole. After this critical value, a fast emergence of the central peak is observed at the cavity frequency for a value of $\tilde{\gamma}_{\theta}$. Finally, the {\it singlet phase} is reached by the system when the DPT has concluded as is shown in Fig.~\ref{fig-1}(g) in the emission spectrum.     
\\
\textit{Conclusions.}\textendash
To the best of our knowledge, this is the first time that a theoretical model within the framework of the Lindblad  master equation explains the intriguing phenomenon of the off-resonant cavity mode emission as a dynamical phase transition, and it being induced by the phonon-mediated coupling. This novel result clarify the existence of the central peak as a signature of the coexistence of the  doublet phase as well as singlet phase in the system, also our numerical results are in perfect agreement with the experimental observations reported by Hennessy {\it et al.} \\
S.E.-A. and H.V.-P. gratefully acknowledge funding by
COLCIENCIAS project  ``Emisi\'on en sistemas de Qubits
Superconductores acoplados a la radiaci\'on. C\'odigo
110171249692, CT 293-2016, HERMES 31361''. S.E.-A. also acknowledges support from the  ``Beca de
Doctorados Nacionales de COLCIENCIAS 727''. E.A.G acknowledges the financial support from Vicerrector\'ia de Investigaciones, Universidad del Quind\'io.


\begin{thebibliography}{99}
\bibitem{Press:2007} D. Press, S. G\"otzinger, S. Reitzenstein, C. Hofmann, A. L\"offler, M. Kamp, A. Forchel, and Yoshihisa Yamamoto, Phys. Rev. Lett. {\bf 98}, 117402 (2007).
\bibitem{Winger:2008} M. Winger, A. Badolato, K.J. Hennessy, E.L. Hu, and A. Imamo\u glu, Phys. Rev. Lett. {\bf 101}, 226808 (2008).
\bibitem{Kaniber:2008} M. Kaniber, A. Laucht, A. Neumann, J. M. Villas-B\^oas, M. Bichler, M.-C. Amann, and J. J. Finley, Phys. Rev. B. {\bf 77}, 161303 (2008).
\bibitem{Hennessy:2007} K. Hennessy, A. Badolato, M. Winger, D. Gerace, M. Atat\"ure, S. Gulde, S. F\"alt, E.L. Hu, and A. Imamo\u glu, Nature {\bf 445}, 896 (2007).
\bibitem{Ota:2009} Y. Ota, N. Kumagai, S. Ohkouchi, M. Shirane, M. Nomura, S. Ishida, S. Iwamoto, S. Yorozu, and Y. Arakawa, Appl. Phys. Express {\bf 2}, 122301 (2009). 
\bibitem{Yamaguchi:2008} M. Yamaguchi, T. Asano, and S. Noda, Opt. Express {\bf 16}, 18067 (2008). 
\bibitem{Naesby:2008}
A. Naesby, T. Suhr, P. T. Kristensen, and J. M\o rk, Phys. Rev. A {\bf 78} 045802, (2008). 
\bibitem{Hughes:2009} S. Hughes, and P. Yao, Opt. Express {\bf 17}, 3322 (2009). 
\bibitem{Auffeves:2009} A. Auffeves, J.-M. Gerard, and J.-P. Poizat, Phys. Rev. A {\bf 79}, 053838 (2009).
\bibitem{Gonzalez:2010} A. Gonzalez-Tudela, E. del Valle, E. Cancellieri, C. Tejedor, D. Sanvitto, F.P. Laussy, Opt. Express {\bf 18}, 7002 (2010). 
\bibitem{Yamaguchi:2009} M. Yamaguchi, T. Asano, K. Kojima, and S. Noda, Phys. Rev. B {\bf 80}, 155326 (2009).
\bibitem{Hohenester:2009} U. Hohenester, A. Laucht, M. Kaniber, N. Hauke, A. Neumann, A. Mohtashami, M. Seliger, M. Bichler, and J.J. Finley, Phys. Rev.
B {\bf 80}, 201311 (2009).
\bibitem{Hohenester:2010} U. Hohenester, Phys. Rev. B {\bf 81}, 155303 (2010).
\bibitem{Hughes:2011}
S. Hughes {\it et al.}, Phys. Rev. B {\bf 83}, 165313 (2011)
\bibitem{Roy:2011b} C. Roy, and S. Hughes, Phys. Rev. Lett. {\bf 106}, 247403 (2011).
\bibitem{Winger:2009}
M. Winger {\it et al.}, Phys. Rev. Lett. {\bf 103}, 207403 (2009).
\bibitem{Chauvin:2009} N. Chauvin, C. Zinoni, M. Francardi, A. Gerardino, L. Balet, B. Alloing, L.H. Li, and A. Fiore, Phys. Rev. B {\bf 80}, 241306 (2009).
\bibitem{Suffczyński:2009} J. Suffczy\'nski, A. Dousse, K. Gauthron, A. Lema\^itre, I. Sagnes, L. Lanco, J. Bloch, P. Voisin, and P. Senellart, Phys. Rev. Lett. {\bf 103}, 027401 (2009).
\bibitem{Majumdar:2010} A. Majumdar, E.D. Kim, Y. Gong, M. Bajcsy, and J. Vu\u ckovi\'c, Phys. Rev. B {\bf 84}, 085309 (2011).
\bibitem{Roy:2011} C. Roy, and S. Hughes, Phys. Rev. X. {\bf 1}, 021009 (2011).
\bibitem{Jarlov:2016} C. Jarlov, \'E. Wodey, A. Lyasota, M. Calic, P. Gallo, B. Dwir, A. Rudra, and E. Kapon, Phys. Rev. Lett. {\bf 117}, 076801 (2016).
\bibitem{Perea:2004} J. I. Perea, D. Porras, and C. Tejedor, Phys. Rev. B. {\bf 70}, 115304 (2004).
\bibitem{Childs:1996} J.J. Childs, K. An, M.S. Otteson, R.R. Dasari, and M.S. Feld, Phys. Rev. Lett. {\bf 77}, 2901 (1996).
\bibitem{Torres:2014} J.M. Torres, Phys. Rev. A. {\bf 89}, 052133 (2014).
\bibitem{Rotter:2010} I. Rotter, J. Mod. Phys. {\bf 1}, 303 (2010).
\bibitem{Jung:1999} C. Jung, M. M\"uller, and I. Rotter, Phys Rev. E. {\bf 60}, 114 (1999).
%
%
%
%
%
%
\end{thebibliography}
\end{document}